\documentclass{article}

\usepackage{graphicx} 
\usepackage{amssymb}
\graphicspath{ {./images/} }
\usepackage[preprint,nonatbib]{neurips_2023}
\usepackage{placeins}

\usepackage[utf8]{inputenc} 
\usepackage[T1]{fontenc}    
\usepackage{hyperref}       
\usepackage{url}            
\usepackage{booktabs}       
\usepackage{amsfonts}       
\usepackage{nicefrac}       
\usepackage{microtype}      
\usepackage{xcolor}         

\usepackage{appendix}

\newcommand{\Lagr}{\mathcal{L}}

\title{Bridging Modalities: Knowledge Distillation and Masked Training for Translating Multi-Modal Emotion Recognition to Uni-Modal, Speech-Only Emotion Recognition} 

\author{Muhammad Muaz \\
  University of Texas at Austin\\
  \texttt{m.muaz@utexas.edu} \\
  \And
   Nathan Paull \\
  University of Texas at Austin\\
  \texttt{napaull@utexas.edu} \\
   \And
   Jahnavi Malagavalli \\
  University of Texas at Austin\\
  \texttt{jahnavimalagavalli@utexas.edu} \\
}

\begin{document}
\maketitle
\date{\today}

\begin{abstract}
    This paper presents an innovative approach to address the challenges of translating multi-modal emotion recognition models to a more practical and resource-efficient uni-modal counterpart, specifically focusing on speech-only emotion recognition. Recognizing emotions from speech signals is a critical task with applications in human-computer interaction, affective computing, and mental health assessment. However, existing state-of-the-art models often rely on multi-modal inputs, incorporating information from multiple sources such as facial expressions and gestures, which may not be readily available or feasible in real-world scenarios. To tackle this issue, we propose a novel framework that leverages knowledge distillation and masked training techniques.
\end{abstract}

\section{Introduction}
Emotion recognition, a pivotal aspect of human-computer interaction and affective computing, has witnessed remarkable progress with the advent of multi-modal models that incorporate information from diverse sources such as facial expressions, gestures, and speech. However, the practical deployment of such sophisticated models is often hindered by the inherent complexity and resource demands associated with processing multiple modalities. In real-world scenarios, obtaining comprehensive multi-modal data may be challenging or impractical, limiting the applicability of these advanced models.

This paper addresses the gap between the theoretical advancements in multi-modal emotion recognition and the practical constraints of real-world applications, with a specific focus on translating these models to a more accessible and efficient uni-modal counterpart—speech-only emotion recognition. Recognizing emotions from speech is crucial in numerous domains, including human-computer interfaces, virtual assistants, and mental health assessment. Our objective is to develop a novel framework that preserves the rich understanding encapsulated in multi-modal models while tailoring it to the constraints and requirements of a speech-only context.

To achieve this, we explore two independent strategies, leveraging knowledge distillation and masked training. Knowledge distillation allows the transfer of nuanced information from a teacher model, originally trained on multi-modal inputs, to a more streamlined student model designed for uni-modal, speech-only emotion recognition. Concurrently, we introduce masked training, a novel training approach designed to reinforce the model's ability to generalize to incomplete inputs, while providing grounding from other modalities. Additionally, we investigate the impact of deep learning-generated audio feature embeddings on the introduced models.

Through a series of experiments, we evaluate the effectiveness of our proposed framework in bridging the gap between multi-modal and uni-modal emotion recognition. The results not only showcase an increased performance of multi-modal models but also demonstrate the competitive capabilities of our proposed models in the challenging domain of speech-only emotion recognition. This work contributes to the ongoing dialogue in the field, emphasizing the need for practical and efficient emotion recognition systems tailored for real-world applications where only speech-related information is readily available.

All code for this work can be found at: \url{https://github.com/m-muaz/Cogmen_SLT}
\section{Related Work}
In the realm of multimodal sentiment analysis, recent research has explored various facets. Addressing cross-modal interaction learning, long-term dependencies in multimodal interactions, and the fusion of unimodal and cross-modal cues, \cite{kumar2020gated} find that learning cross-modal interactions significantly benefits the task. Their experiments on CMU Multimodal Opinion level Sentiment Intensity (CMU-MOSI) and CMU Multimodal Opinion Sentiment and Emotion Intensity (CMU-MOSEI) datasets demonstrate accuracies of 83.9\% and 81.1\%, respectively, showcasing a noteworthy improvement over the current state-of-the-art. Another approach, presented by \cite{Datcu}, focuses on bimodal emotion recognition using face and speech analysis, leveraging Hidden Markov Models (HMMs) for temporal dynamics and introducing novel methods for feature selection. Additionally, \cite{delbrouck-transformer} contribute a Transformer-based joint-encoding (TBJE) for Emotion Recognition and Sentiment Analysis, incorporating modular co-attention and a glimpse layer for multimodal encoding. \cite{lian20c_interspeech} propose a graph-based neural network for conversational emotion analysis, modeling context-sensitive and speaker-sensitive dependences, with an absolute improvement of 1\%--2\% over state-of-the-art strategies demonstrated on the MELD dataset. Lastly, \cite{cogmen} introduces COGMEN, a Contextualized Graph Neural Network for Multimodal Emotion Recognition, leveraging local and global information in conversations to achieve state-of-the-art results on IEMOCAP and MOSEI datasets. These works collectively showcase diverse strategies and advancements in multimodal sentiment analysis and emotion recognition, emphasizing the importance of cross-modal interactions, dynamic modeling, joint-encoding, and graph-based approaches.

\section{Approach}
\subsection{Dataset}
 In this work, we leverage the IEMOCAP4 dataset, a rich and diverse corpus that encompasses speech, video, and text features, facilitating a holistic understanding of emotional expressions. The dataset is specifically curated to recognize four distinct emotion classes: neutral, happy, sad, and angry.

Our focus within the IEMOCAP4 dataset is on exploiting the speech and video modalities, mirroring the sensory perception of human emotions. Restricting our analysis to these modalities aligns with the practical constraints of real-world applications, where obtaining text-based features may not always be feasible. The decision to utilize speech and video features allows us to create a more applicable and resource-efficient model for scenarios where only these modalities are readily available.
\subsection{Model Architecture}
\begin{figure}[htb!]
    \centering
    \includegraphics[width=1\linewidth]{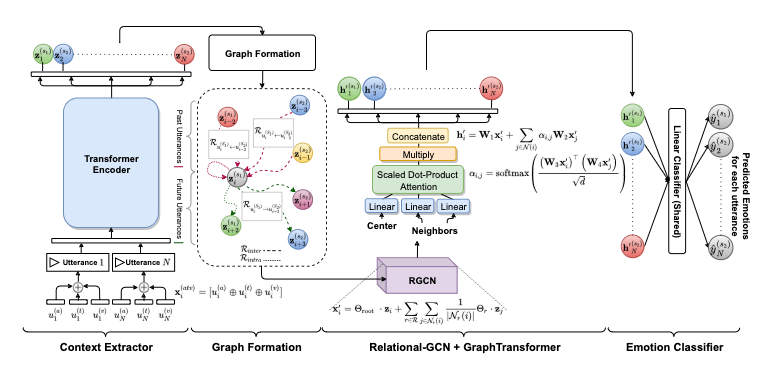}
    \caption{COGMEN Model Architecture \cite{cogmen}}
    \label{fig:cogmen}
\end{figure}

To benchmark the performance of our proposed framework for translating multi-modal emotion recognition to uni-modal, speech-only emotion recognition, we integrate the state-of-the-art COGMEN architecture as shown in Figure \ref{fig:cogmen}. The COGMEN architecture stands as the pinnacle in multi-modal emotion recognition on the IEMOCAP4 dataset, incorporating three key components that capture and exploit the inter-modal dependencies inherent in emotional expressions. 

\FloatBarrier
\FloatBarrier

The first component is a transformer-based feature extractor used to capture complex intra-utterance dependencies. This component is instrumental in extracting high-level representations from the input data, enabling the model to discern nuanced patterns associated with different emotional states.

The second component is a GNN, where utterances are connected to their \textbf{i}-precursor and \textbf{j}-successor utterances in time, forming a temporal graph that encapsulates the dynamic evolution of emotions over a length of conversation. This enables the model to leverage the temporal context and relationships between utterances, facilitating a more holistic understanding of emotional dynamics. In our experiments and in the original COGMEN paper \textbf{i} and \textbf{j} are both set to 5.
\subsection{Feature Embeddings}
COGMEN utilizes the OpenSmile toolkit \cite{Eyben_openSMILE} to extract audio features (size 100). This toolkit integrates algorithms from speech processing and Music Information Retrieval, employing classic feature extractors such as the Mel-band filter-bank. In our project, we investigate the impact of employing deep learning feature extractors like HuBERT \cite{hsu2021hubert} on the predictions of the base COGMEN, Masked COGMEN, and Knowledge Distillation COGMEN models. HuBERT features are extracted from its second-to-last layer, preceding the classification layer (size 1024 after applying mean over dim = 1). Given the input audio feature size of 100, a linear projection is applied to reduce the HuBERT output feature dimension to 100. Despite these generated features, the models exhibited poor performance, as detailed in the appendix. A potential reason for this may be the linear projection layer altering the feature values. Further exploration is needed to draw conclusive results.

\subsection{Knowledge Distillation}

The SOTA performance of COGMEN \cite{cogmen} architecture showcases that multi-modality indeed outperforms uni-modal approaches which is evident from the fact that in nature humans and other living beings utilize information received from multiple senses. However, incorporating multiple inputs comes with additional computational cost. Even when resources are abundant, having resource-efficient networks means more requests can be served at a lower cost. Along these lines, we try to answer the question: \textit{Can we get a speech-only emotion recognition model while still maintaining the performance of the larger and efficient model?} To answer this, we focused along the domain of knowledge distillation \cite{hinton2015distilling}. Knowledge distillation is a general technique for supervising the training of "student" neural networks by capturing and transferring the knowledge of trained "teacher" networks (In our context, "teacher" model refers to audio-visual input COGMEN model and "student" model refers to audio only COGMEN model). The trained teacher network provides additional semantic knowledge besides the usual data supervision. However, the question of how to encoder and utilize the teacher's knowledge such that student's performance is maximized. 

For this, we hypothesize that supervision of multiple modalities (audio, video, text, etc.) may have generated more rich embedding distributions/representations \footnote{We use the embeddings after the Relational Graph Convolutional Network (RCGN)+Graph Transformer module}. So, it would be beneficial if we can use that rich embedding space to guide the student network. In order to generate the supervision signal for the student model, we utilize triplet loss. Formally, let $X$, $P$ and $N$ be the student model embedding representation , teacher model embedding representation for positive sample, and teacher embedding for negative sample, respectively. The loss function can then be described by means of p-norm:
\[
\Lagr_{triplet}(X,P,N) = \max(\|f(X) - g(P)\|_{p} - \| f(X) - g(N) \|_{p} + \alpha, 0)
\]
where $\alpha$ is a margin between positive and negative pairs and f and g be the student and teacher embedding networks. For our experiments, $\alpha$ has been set to $1$ and $p$ to $2$. 

In addition to the signal from multi-class cross entropy signal to force the model to accurately predict emotions, the triplet loss will generate the guidance signal for the student to mimic the embeddings learnt by the teacher network. Therefore, the student model is trained to optimize the following multi-term loss:
\[
    \Lagr = \alpha_{1} \Lagr_{triplet} + \alpha_{2} \Lagr_{MC-CE}
\]
where MC-CE refers to multi-class cross entropy loss. For our experiments, we kept both alpha parameters to 1 and it seems to be working for our experiments. However, future works could explore the impact of different parameters values either via manual value setting or making these parameters learnable.

\subsection{Masked Training}
In the pursuit of adapting the COGMEN architecture to a uni-modal, speech-only emotion recognition context, our proposed framework incorporates a novel training strategy known as masked training. This technique aims to refine the model's attention mechanisms specifically for speech-related features while maintaining performance on multiple modalities. Our masked training routine stochastically chooses between four distinct masking scenarios. 

\begin{figure}[htb]
    \centering
    \includegraphics[width=0.85\linewidth]{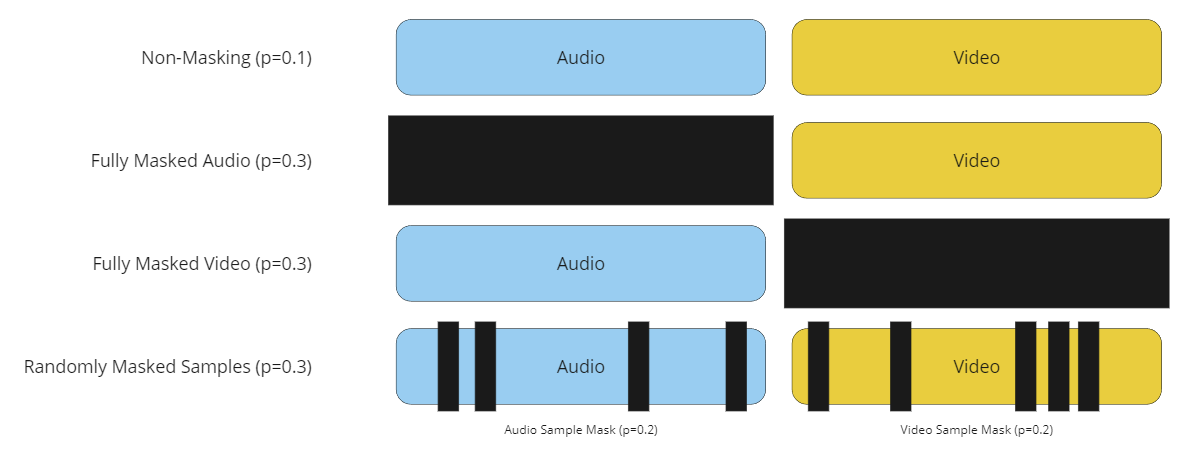}
    \caption{An illustration of the 4 input masking scenarios and their probabilities used during training.}
    \label{fig:mmask-demo}
\end{figure}

As shown in the figure above these four masking scenarios are non-masking, fully masked audio, fully masked video, and randomly masked samples. In our exploration we found that choosing non-masking with a probability of 0.1, fully masking audio or video each with a probability of 0.3, and then with a probability of 0.3, choosing to randomly mask samples. In the case of randomly masking samples, each sample is chosen as the start of a mask with probability of 0.2 and the masks are of length 1 for both audio and video. It is worth noting that the probability of choosing the random sample masking is the probability of choosing this masking strategy, which is different from the probability of masking each sample if this masking strategy is chosen. 

This method provides a large number of hyper-parameters to explore such as the probability of choosing each masking method and the length of sample masks for each modality. While we believe we have found a high performing set of hyper-parameters, we do not exclude the possibility that other hyper-parameters made lead to better results, however, most other hyper-parameter choices that we explored led to worse results.

\section{Results}
\subsection{Knowledge Distillation}

Our teacher model and student model will be referred to as COGMEN$_{T}$-AV and COGMEN$_S$-A respectively to indicate that teacher model is the multi-modal COGMEN model while student model is the uni-modal network.
\FloatBarrier
\begin{table}[htbp]
    \centering
    \caption{F1 Scores on the Test Data}
    \label{table:kd-f1}

    \begin{tabular}{l c }
        \toprule
        \textbf{COGMEN Models} & \textbf{F1-scores (Inference)} \\
        \midrule
        COGMEN$_T$-AV [SeqContext:4, GNN:7]& $0.6099_{\pm 0.0197}$ \\
        COGMEN$_S$-A [SeqContext:4, GNN:7]& $0.5999_{\pm 0.0276}$ \\
        \midrule
        COGMEN$_T$-AV [SeqContext:4, GNN:7]& $0.6440_{\pm 0.02}$ \\
        COGMEN$_S$-A [SeqContext:\textbf{2}, GNN:7]& $0.6038_{\pm 0.0286}$ \\
        \midrule
        COGMEN$_T$-AV [SeqContext:4, GNN:$7$]& $0.6112_{\pm 0.0072}$ \\
        COGMEN$_S$-A [SeqContext:4, GNN:\textbf{3}]& $0.5857_{\pm 0.0426}$ \\
        \bottomrule
    \end{tabular}
\end{table}
\FloatBarrier

\begin{figure}[htb]
    \centering
    \includegraphics[width=0.85\linewidth]{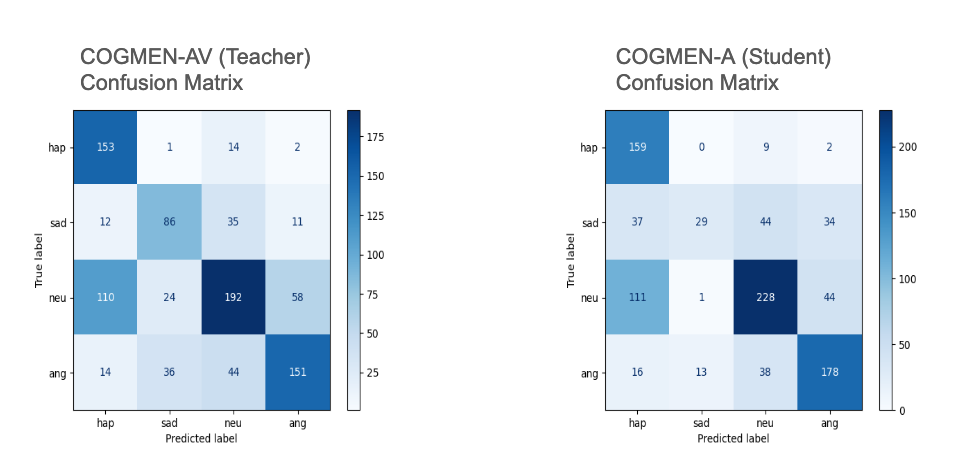}
    \caption{Comparison of the confusion matrix for the teacher and student models.}
    \label{fig:kd-cm}
\end{figure}

The above table \ref{table:kd-f1} shows the F1 metric scores of both teacher and student models during inference. SeqContext means the number of transformer layers used in the Context Extractor module of the COGMEN model while, GNN refers to the number of transformer layers in the GraphTransformer module of COGMEN.
The first row shows that if both teacher and student model are using the same architecture then, the student can achieve similar performance while only relying on speech input modality. Moreover, as show in the Figure \ref{fig:kd-cm} , it can be seen that the distribution learnt by student model closely resembles the distribution of its teacher. Among all the emotions, the student model is more uncertain during the prediction of \textbf{sad} emotion.

The second corresponds to the ablation experiment aimed to analyze whether or not the student model show robustness when number of transformer layers, responsible for extracting contextual information, are reduced.
Moreover, the third row corresponds to the experiment focused to analyze the impact of reducing number of transformer layers in the graph transformer module. From the results, it can clearly seen that reducing the number of transformer layers in the context extractor (Initial stage) of student model have a significant impact as compared to the variation in number of layer in the Graph Transformer (later stage). However, the student model is pretty robust against both variations as we only see a marginal drop of $\approx 0.04$ and $\approx 0.02$. These results also show that the initial stage of context extractor plays a significant role in model' performance which is intuitive as if the model extracts good contextual representations from input then it can help the model generate better representations during the intermediate and later stage of the model.
\FloatBarrier
\FloatBarrier

All in all, it can be concluded that the uni-modal student model effectively achieves the same performance as of its teacher while also showing robustness to the variation in different aspects of the underlying model.
\subsection{Masked Training}
Our masked training adjusted COGMEN model will be called COGMEN-Mask, which we will compare to an audio-only COGMEN model (COGMEN-A) and an audio+video COGMEN model that will be tested in an audio-only context (COGMEN-AV).

\FloatBarrier
\begin{table}[htb]
    \centering
    \caption{Results of the masked-training routine compared to baseline models with no training adjustments.}
    \begin{tabular}{c | c | c}
       Models  & F1 Score (AV Inference) & F1 Score (A Inference) \\
       \hline \hline
       COGMEN-A \cite{cogmen} & - & 0.636 \\
       COGMEN-AV \cite{cogmen}  & 0.645 & 0.426 \\
       COGMEN-Mask (Ours)  & \textbf{0.687} & \textbf{0.662} \\ 
    \end{tabular}
    
    \label{tab:mmask_f1}
\end{table}
\FloatBarrier

As shown, the COGMEN-AV model sees a large decrease in performance when directly applied to the audio-only context with an F1 decrease of over 0.2 and an audio-only performance far lower than the COGMEN-A model. Additionally, we see that our model, COGMEN-Mask, not only is able to mitigate this F1 loss, seeing a performance decrease of approximately 0.02, but is able to outperform both the COGMEN-AV and COGMEN-A models in both inference contexts. With additional adjustments to the COGMEN model regarding the number of sequential layers and GNN heads, we have seen audio-only performance increase up to 0.674, however, this gave an audio-video score of 0.622, performing worse when given the additional video modality. We chose not to incorporate this score in the table as audio-only outperforming audio-video implies that the model is not generalizing as well as the model recorded in the table. This loss of generalization appears to be caused by an increase in the number of sequential layers from the default of two.

\FloatBarrier
\begin{figure}[htb]
    \centering
    \includegraphics[width=0.9\linewidth]{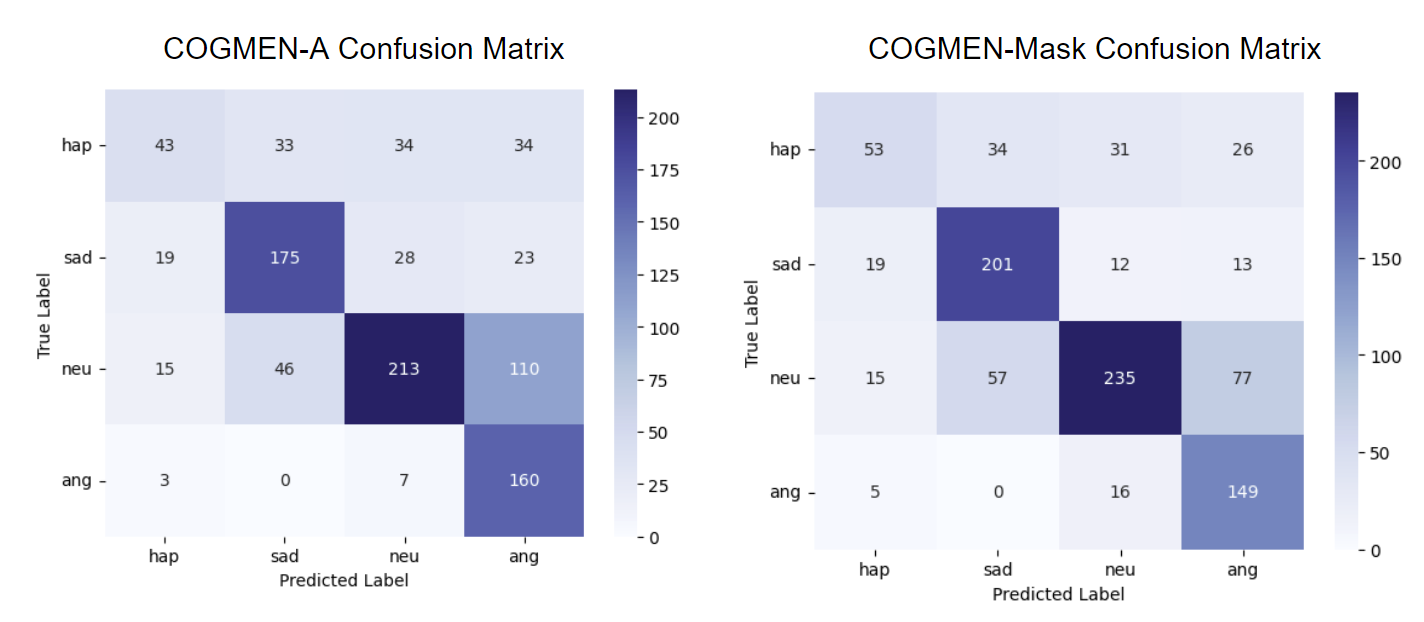}
    \caption{Comparison of the confusion matrix for the COGMEN-A and COGMEN-Mask models.}
    \label{fig:mmask-cm}
\end{figure}
\FloatBarrier

Additionally, when comparing the performance on individual emotion classes for the COGMEN-A and COGMEN-Mask models during audio-only inference, we don't see the COGMEN-Mask model achieve vastly different performance on any single emotion class. Instead, the model appears to learn slightly more robust representations for all emotion classes. 

Because COGMEN-Mask was more effective in audio-video emotion recognition we chose to explore using COGMEN-Mask as our teacher model in a knowledge distillation training setup. The student model from this setup was able to achieve an audio-only score of \textbf{0.650}, which while less that the COGMEN-Mask model, implies that the knowledge distillation setup is likely to benefit from any performance gains found in the teacher model.

The last question regarding masked training that we seek to answer in this paper is whether the COGMEN-A model is truly an adequate baseline for the COGMEN-Mask model, i.e. would a uni-modal version of our masked training routine create similar benefits for the COGMEN-A model. Our observations when running a uni-modal adaptation of the masked training on the COGMEN-A model yielded better results than the default COGMEN-A model, with the best F1 score we saw being \textbf{0.654}, however, this still fails to exceed the performance of the COGMEN-Mask model and required adjustments to the number of sequential layers and GNN heads in the COGMEN model, thus requiring greater hyper-parameter search. 

Overall, these results show that COGMEN-Mask not only achieves better translation from multi-modal to uni-modal contexts, but additionally seems to gain some knowledge grounding by being trained in a multi-modal context. Additionally, we ran experiments to discover how valuable COGMEN's GNN module is to the observed masked training performance increase. Our findings indicate that random input masking is able to provide performance increases in both audio-only and audio-video contexts without the inclusion of a GNN, suggesting that this training method will generalize across model architectures. For full results see: Table \ref{tab:no_gnn}.

\section{Conclusion}

We found that both knowledge distillation and masked training are effective avenues for preventing performance drop-off when translating from multi-modal to speech-only emotion recognition. The masked training not only mitigated performance drop-off but even improved over both the audio-video and audio-only benchmarks, this came at the cost of greater hyper-parameter tuning. This hyper-parameter space will continue to increase as the number of modalities increases which may make this method intractable for models with a high number of modalities or long training times. The knowledge distillation approach does not have this modality dependent hyper-parameter growth and instead only adds two hyper-parameters to control the balance between predictive and contrastive loss, which may ultimately make it the more desirable option between the two approaches discussed in this paper.

\section{Future Work}
The baseline COGMEN model that we have built our work on top of mainly achieves its SOTA multi-modal emotion recognition performance from the text modality as compared to the speech or vision modalities. This is rather interesting as humans often experience the opposite, struggling to recognize emotion in a text-only modality while excelling in any context where speech and/or vision are present. The audio features used in our most successful experiments came from OpenSmile with features coming from learned feature extractors such as HuBERT and other Transformers showing a decrease in performance. It is possible that SSMs \cite{s4,ls4,s5}, being designed to learn on high frame-rate/continuous data, may provide a reduction in the human-machine modality preference gap.

Altering the COGMEN model so that emotion recognition is done in a sequential or even real-time manner is another place for future research. The current graph used by the GNN uses both past and future utterances for emotion extraction. Limiting all edges in the graph such that an utterance can only predict using itself an prior utterances would bring this model closer to real-time emotion recognition. Exploration of masked training and knowledge distillation in this domain may similarly increase performance of these models. Similarly, using a knowledge distillation setup, one could use a bi-directional COGMEN teacher to instruct a sequential COGMEN student.

The COGMEN model is an intermediate fusion model, performing independent feature embeddings for each modality before they are fused in the transformer layer. The level at which this fusion occurs should not aggressively affect the knowledge distillation approach, however, there is a chance that masked training adjusted for early or late modality fusion may see results that are quite different. 


\newpage
{\small
\bibliographystyle{unsrt}
\bibliography{main}
}

\newpage
\appendix
\appendixpage 

\section{Input Masked Emotion Recognition without a GNN}
At the end of our COGMEN-Mask experiments we were left with a lingering question, "How important is the GNN to the performance of the COGMEN-Mask model?" Because the input masking is stochastic, it is unlikely that an utterance will share the same input mask as its neighboring utterances in the past and future, as such it is possible that the GNN is vital to the performance gain from random input masking. This meant that experiments that removed the GNN component would be key to understanding the effectiveness of this input masking method across varied underlying models. 

Without using a completely new model, we decided to use the COGMEN model, but adjusted the utterance graph to be completely disjoint, such that all utterances were only connected to themselves. This makes the COGMEN GNN a simple feed-forward layer that does not rely on any external information. After making this adjustment we re-ran the masking experiments and listed our results in the table below.

\FloatBarrier
\begin{table}[htb]
    \centering
    \caption{Results of the masked-training routine with a disjoint utterance graph.}
    \begin{tabular}{c | c | c}
       Models  & F1 Score (AV Inference) & F1 Score (A Inference) \\
       \hline \hline
       COGMEN-A \cite{cogmen} & - & 0.558 \\
       COGMEN-AV \cite{cogmen}  & 0.367 & 0.286 \\
       COGMEN-Mask (Ours)  & \textbf{0.604} & \textbf{0.607} \\ 
       COGMEN-A + Mask & - & 0.571
    \end{tabular}
    
    \label{tab:no_gnn}
\end{table}
\FloatBarrier

These experiments clearly showed how the masked input method is highly performant even without the underlying GNN, thus implying that it is likely to improve performance across all model classes. It is worth noting that the COGMEN-AV results appear to be an outlier compared to all other results. We are unsure why this is, however, it is possible that the extra modalities provide too much noise to the learning process and without normalization from a GNN or randomly masked input, the model will fail to converge.

\section{Feature Embeddings Results and Analysis}
The HuBERT-generated audio features passed to the base, Masked, and Knowledge Distillation COGMEN models resulted in the following F1 scores. Table \ref{basemodelf1} presents the results of the original COGMEN model when trained and tested with 'a' and 'av' modalities, respectively, using COGMEN-generated audio features and HuBERT audio features. Please note that some data points had to be removed from the COGMEN features as they were not present in Kaggle's IECOMAP dataset. This reduction in data points might be another reason for the reduction in the F2 scores compared to the COGMEN A and COGMEN AV models results in Table \ref{tab:mmask_f1}.

\begin{table}[htbp]
    \centering
    \caption{Base model F1 scores on the Test Data}
    \label{basemodelf1}
    \begin{tabular}{lcc}
        \toprule
        \textbf{COGMEN Models} & \textbf{A inference} & \textbf{AV inference} \\
        \midrule
        \textbf{Base - COGMEN} & 0.56 & 0.56 \\
        \textbf{Base - HuBERT} & 0.52 & 0.53 \\
        \bottomrule
    \end{tabular}
\end{table}

Table \ref{ouremodelsf1} presents the F1 scores of the masked and distilled training COGMEN models. In both Tables \ref{basemodelf1} and \ref{ouremodelsf1}, the F1 scores of the models trained on COGMEN audio features are higher than those of HuBERT audio features. This might be because the HuBERT features were passed to a linear projection layer, which changes the actual feature values.
\begin{table}[htbp]
    \centering
    \caption{F1 scores on the Test Data}
    \label{ouremodelsf1}

    \begin{tabular}{lcc}
        \toprule
        \textbf{COGMEN Models} & \textbf{A inference} & \textbf{AV inference} \\
        \midrule
        \textbf{Masked - COGMEN AV} & 0.57 & - \\
        \textbf{Masked - HuBERT AV} & 0.2282 & -  \\
        \midrule
        \textbf{Student - COGMEN A} & 0.5964 & - \\
        \textbf{Student - HuBERT A} & 0.3738 & - \\
        \midrule
        \textbf{Teacher - COGMEN AV} & - & 0.6369  \\
        \textbf{Teacher - HuBERT AV} & - & 0.55675\\
        \bottomrule
    \end{tabular}
\end{table}

\begin{figure}[htb]
    \centering
    \includegraphics[width=0.9\linewidth]{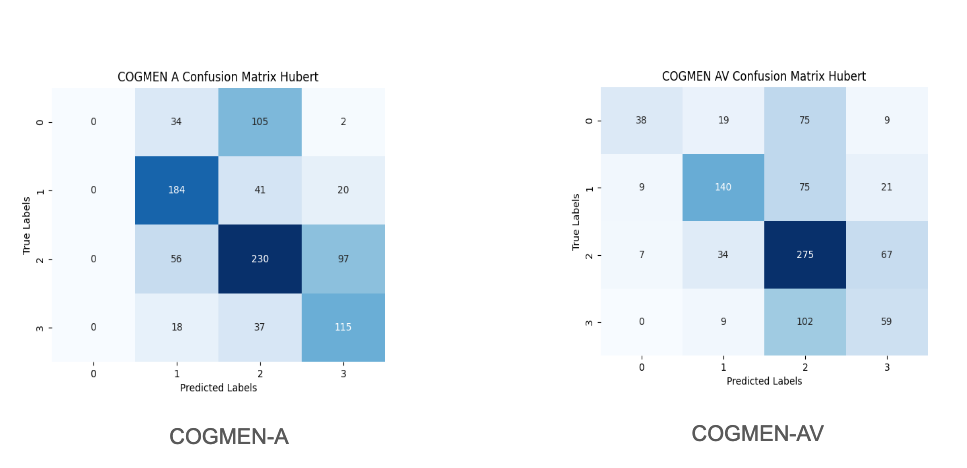}
    \caption{Confusion matrices with HuBERT audio features on Base COGMEN}
    \label{fig:COGMEN}
\end{figure}

\begin{figure}[htb]
    \centering
    \includegraphics[width=0.9\linewidth]{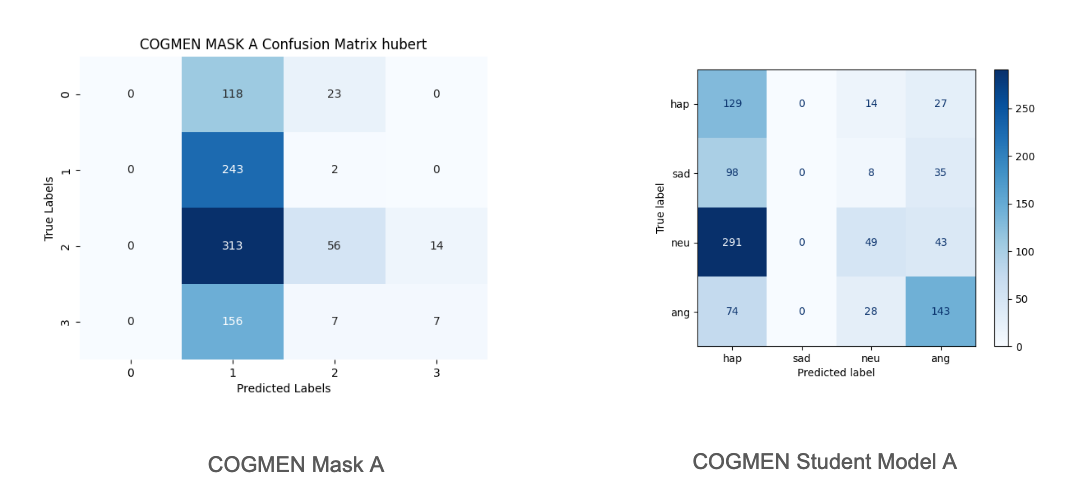}
    \caption{Confusion matrices of models with HuBERT audio features.}
    \label{fig:COGMEN_Modified}
\end{figure}

Another reason for these poor results might be the hyperparameter settings. The hyperparameters used were the same as those used in each of the three models described in the paper. Confusion matrices in Figures \ref{fig:COGMEN} and \ref{fig:COGMEN_Modified} for these models with HuBERT features on the test dataset are not good due to the aforementioned reasons.

\FloatBarrier
\FloatBarrier

\FloatBarrier
\FloatBarrier

     


\end{document}